# NUCLEAR PHOTONICS AT ULTRA-HIGH COUNTING RATES AND HIGHER MULTIPOLE EXCITATIONS

P.G. Thirolf[a], D. Habs[a,b], D. Filipescu[c], R.Gernhäuser[d], M.M. Günther[b], M. Jentschel[e], N. Marginean[d], N. Pietralla[f]

[a]*Fakultät f. Physik, Ludwig-Maximilians-Universität München, Garching, Germany*
[b]*Max-Planck-Institute f. Quantum Optics, Garching,Germany*
[c]*IFIN-HH, Bucharest-Magurele, Romania*
[d]*Physik Department E12,Technische Universität München, Garching, Germany*
[e]*Institut Laue-Langevin, Grenoble, France*
[f]*Institut f. Kernphysik, Technische Universität Darmstadt, Germany,*

**Abstract.** Next-generation γ beams from laser Compton-backscattering facilities like ELI-NP (Bucharest)] or MEGa-Ray (Livermore) will drastically exceed the photon flux presently available at existing facilities, reaching or even exceeding $10^{13}$ γ/sec. The beam structure as presently foreseen for MEGa-Ray and ELI-NP builds upon a structure of macro-pulses (~120 Hz) for the electron beam, accelerated with X-band technology at 11.5 GHz, resulting in a micro structure of 87 ps distance between the electron pulses acting as mirrors for a counterpropagating intense laser. In total each 8.3 ms a γ pulse series with a duration of about 100 ns will impinge on the target, resulting in an instantaneous photon flux of about $10^{18}$ γ/s, thus introducing major challenges in view of pile-up. Novel γ optics will be applied to monochromatize the γ beam to ultimately $\Delta E/E \sim 10^{-6}$. Thus level-selective spectroscopy of higher multipole excitations will become accessible with good contrast for the first time. Fast responding γ detectors, e.g. based on advanced scintillator technology (e.g. $LaBr_3(Ce)$) allow for measurements with count rates as high as $10^6$-$10^7$ γ/s without significant drop of performance. Data handling adapted to the beam conditions could be performed by fast digitizing electronics, able to sample data traces during the micro-pulse duration, while the subsequent macro-pulse gap of ca. 8 ms leaves ample time for data readout. A ball of $LaBr_3$ detectors with digital readout appears to best suited for this novel type of nuclear photonics at ultra-high counting rates.

## INTENSE, BRILLIANT γ BEAMS FROM LASER COMPTON BACKSCATTERING

Within the next few years γ beams from upcoming next-generation laser Compton-backscattering facilities like MEGa-Ray (Livermore, operation planned for 2013) [1] and ELI-NP (Bucharest, envisaged start of operation end of 2016) [2] hold promise of exceeding the photon flux presently available at existing facilities like HIγS (Duke University) [3] by several orders of magnitude, reaching or even exceeding $10^{13}$ γ/s. Based on a linear electron accelerator, the Doppler-upshift of Compton-backscattered laser photons off relativistic electrons is exploited to generate a high-energy (up to 19 MeV envisaged for ELI-NP) and highly brilliant γ beam for basic photonuclear physics and its applications. For ELI-NP an average photon flux of $10^{13}$ γ/s is envisaged, which due to the low duty-cycle of ~$10^{-5}$ results in a huge instantaneous rate of $10^{18}$ γ/s, way beyond the capabilities of any photon detection system. Thus 'tailoring' an optimized γ beam experiment from the γ beam generation over an efficient monochromatization up to a γ detection system with highest count-rate capabilities is an indispensable necessity. Refractive lenses, so far routinely used at synchrotron facilities for focusing γ energies up to 200 keV [4], are presently studied to evaluate their potential also for higher γ energies up to the MeV range [5,6]. If successful, a crystal monochromator comparable to the GAMS device installed at the ILL [5,8], however significantly boosted in efficiency [6], could be applied to reach an ultimate energy band-width of $\Delta E/E=10^{-6}$, corresponding to an energy resolution of only a few eV. The resulting photon flux can still be kept as high as $10^7$ γ/s on average, corresponding to an instantaneous rate of $10^{12}$ γ/s and equivalent to a spectral flux of ca. $10^6$-$10^7$ /(eV·s) (to be compared to the presently most intense γ source HIγS at Duke University with a spectral flux (at $\Delta E/E$~3%) of $10^2$/(eV·s)). It is obvious that only a γ detection system with highest count-rate capability will be able to operate under such severe conditions.

## γ BEAM PULSE STRUCTURE

The beam pulse structure as presently foreseen for MEGa-Ray and ELI-NP builds upon a structure of macro-pulses (~120 Hz) for the electron beam, accelerated with X-band technology at 11.5 GHz, resulting in a micro structure of 87 ps (≈2.6 cm) distance between the (ca. 1000) electron bunches (each of them ca. 0.5 ps long with an electron charge of about 25 pC) acting as mirror for a counter-propagating intense laser with a long pulse length of ~10 ns [7]. After interaction with an electron bunch, the laser pulse is recirculated ca. 8 times in an optical cavity (ca. 1 m long, thus simultaneously containing ca. 38 electron bunches). In total each 8.3 ms a series of 0.5 ps γ pulses with a duration of about 100 ns will impinge on the target. In view of the envisaged integral photon flux of $10^{13}$ γ/s, this scheme assures that very efficiently 2/3 of all electrons contribute to the γ beam production. Presently the macro-pulse frequency is limited by the performance of the expensive customized RF clystrons. Drawing on an envisaged industrial mass production of such devices driven by the needs of future high-energy accelerator projects, the γ beam duty cycle could be further increased.

## HIGH COUNT-RATE PHOTON DETECTION SYSTEM

Pile-up has to be considered as a serious limitation, forcing to use fast-responding γ detection systems with high count- rate capabilities and/or digital readout and post-processing concepts. Already today, commercially available fast digitizing systems provide a digitizing rate of 5 GSample/s with an energy resolution of 12 bit and 1024 storage cells per channel [9]. This allows for digitizing the signal trace of a full micro pulse. The subsequent 8.3 ms interval to the next macro pulse leaves ample time for readout, even in view of the high deadtime of 100 μs for such systems due to their switched-capacitor technology.

Novel, extremely fast scintillator materials like LaBr$_3$(Ce) have already demonstrated high-count rate capabilities of about 1 MHz without further pile-up treatment, while advanced pile-up correction algorithms and digital filters could push this limit to about 10 MHz [10]. Such detectors may even be fast enough to resolve the individual micro pulse structure, allowing for localizing the temporal γ origin to ~0.5 ps. Tab. 1 displays typical detector properties of LaBr$_3$ scintillators compared to germanium detectors (for the same active volume of 3x3").

**TABLE 1.** Typical properties of LaBr$_3$ scintillation detectors compared to germanium solid state detectors of the same active volume (3x3" in each case).

|  | Scintillator (LaBr$_3$) | Germanium |
|---|---|---|
| Energy range | keV – MeV | keV – MeV |
| Abs. efficiency (d=1") | $10^{-1} – 10^{-2}$ | $2·10^{-1} – 5·10^{-2}$ |
| Rel. efficiency (1.33 MeV) | 143% | 75% |
| Energy resolution (1 MeV) | 20-30 keV | 1-2 keV |
| Time resolution | 200 ps (@ 511 keV) | 20 ns – 5 ns |

Thus the optimum γ spectroscopy setup for ELI-NP could consist of a ball of 10-20 LaBr$_3$ crystals, covering a solid angle of about 10%. Small targets even from rare isotopes can be used due to the well-focused γ beam.

# LEVEL-SELECTIVE γ SPECTROSCOPY AND HIGHER MULTIPOLE EXCITATIONS

So far photonuclear measurements can be at best performed with an energy resolution of about 3% or ΔE~100 keV, while older experiments had to deal with the full bremsstrahlung spectrum. Thus dominant E1 or M1 excitations hampered the observation of weak higher multipole excitations in their vicinity, except for the well-established use of nuclear resonance fluorescence for studies of E2 excitations [11]. In contrast, given the high spectral flux with an energy bandwidth only limited by the thermal Doppler level broadening of typically 1-10 eV, γ spectroscopy of individual levels will become accessible at ELI-NP and MEGa-Ray. Previously unobservable higher multipole excitations could be excited for the first time with good contrast and few-eV resolution without being disturbed by a neighbouring E1 or M1 resonance. In view of the much smaller intrinsic widths of these higher multipoles, the very intense γ beams of MEGa-Ray and ELI-NP will prove beneficial. Each of those higher multipoles will decay via a characteristic decay pattern, e.g. a $2^+$ core excitation will decay via E2 transitions to $0^+$, $2^+$ or $4^+$ states. In case of such rather simple decay patterns, the energy resolution provided by LaBr$_3$ detectors will be sufficient to resolve these structures, while the relative γ beam energy resolution of up to $10^{-6}$ will allow for localizing the resonances to few eV. Moreover, due to the adjustment of the beam energy to individual resonances, otherwise strong atomic background from

Compton scattering or pair creation will be drastically reduced. For polarized γ beams angular correlations will allow for extracting the parity information. A variety of elementary collective excitations could be studied: quadrupole shape vibrations (β, γ band), double scissors mode, rotational states built on the scissors mode (SM), or multiphonon quadrupole $Q^n$ resonances (e.g. double octupole phonon in $^{96}$Zr/$^{208}$Pb). The M2 twist mode (e.g. in $^{90}$Zr), spherical and deformed octupole vibrations, mixed-symmetry octupole states or octupole-skin vibrations as well as (E4) hexadecupole vibrations and Pygmy resonances (PR) could be targeted. Thus photonuclear physics would not be dominated any longer by E1 and M1 excitations and it will be interesting to study to what extent these higher multipolarities will become accessible. The strength distribution of the individual multipole resonances could be studied in detail, providing information also on level densities and damping widths.

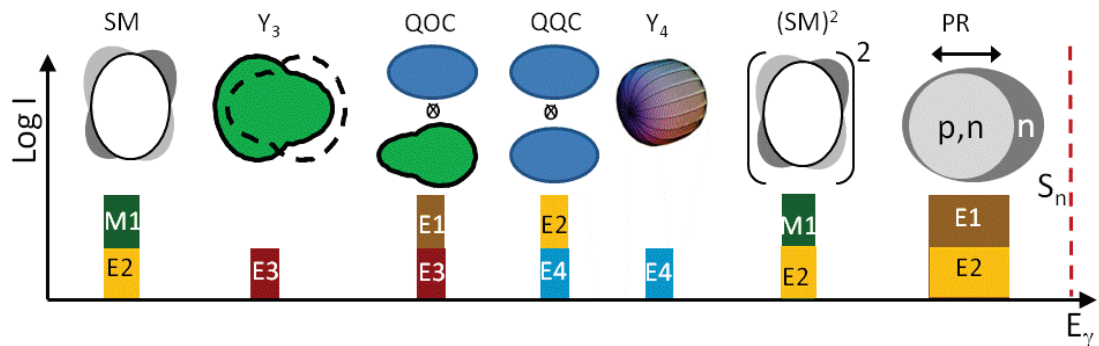

**FIGURE 1.** Schematic picture of multipole excitations and their decay patterns.

From a practical point of view, scanning the γ excitation energy when searching for new resonances will require a re-adjustment of the maximum tolerable count rate in the detector ball for each resonance. This could be achieved e.g. by providing a selection of targets with different thicknesses.

In conclusion, the unprecedented properties of brilliant γ beams soon available at ELI-NP and MEGa-Ray will open up intriguing new perspectives for level-selective γ spectroscopy of higher multipole excitations, based on new concepts of γ optics, fast detectors and digital readout to digest the expected ultra-high counting rates.